\newcommand{\CLASSINPUTtoptextmargin}{50pt}
\newcommand{\CLASSINPUTbottomtextmargin}{50pt}
\newcommand{\CLASSINPUTinnersidemargin}{41pt}
\newcommand{\CLASSINPUToutersidemargin}{41pt}

\documentclass[9pt, sigconf, nonacm, screen=true,bookmarks=false]{acmart}

\usepackage[utf8]{inputenc} %
\usepackage[T1]{fontenc}    %
\usepackage{pifont}

\usepackage{algorithm}
\usepackage{graphicx}
\usepackage{threeparttable}
\usepackage{multirow}
\usepackage{amsmath}
\usepackage{bm}
\usepackage{bbm}
\usepackage{amsthm}
\usepackage{soul}
\usepackage[noend]{algpseudocode}
\usepackage{enumitem} %
\usepackage[subrefformat=parens,labelformat=parens]{subfig}

\usepackage{wrapfig}
\usepackage{xspace}

\definecolor{citecolor}{RGB}{34,139,34}
\definecolor{mydarkblue}{rgb}{0,0.08,1}
\definecolor{mydarkgreen}{rgb}{0.02,0.6,0.02}
\definecolor{mydarkred}{rgb}{0.8,0.02,0.02}
\definecolor{mydarkorange}{rgb}{0.40,0.2,0.02}
\definecolor{mypurple}{RGB}{111,0,255}
\definecolor{myred}{rgb}{1.0,0.0,0.0}
\definecolor{mygold}{rgb}{0.75,0.6,0.12}
\definecolor{myblue}{rgb}{0,0.2,0.8}
\definecolor{mydarkgray}{rgb}{0.,0.2,0.2}

\definecolor{lightred}{RGB}{255,235,235}
\definecolor{lightgreen}{RGB}{235,255,235}
\definecolor{lightblue}{RGB}{235,235,255}
\definecolor{lightcyan}{RGB}{235,255,255}
\definecolor{lightmagenta}{RGB}{255,235,255}
\definecolor{lightyellow}{RGB}{255,255,235}

\definecolor{qxkcolor}{RGB}{215,235,255}
\definecolor{softmaxcolor}{RGB}{230,235,255}
\definecolor{probxvcolor}{RGB}{255,255,235}

\definecolor{topkcolor}{RGB}{255,235,235}
\definecolor{zecolor}{RGB}{255,255,235}
\definecolor{dynacolor}{RGB}{235,255,255}

\definecolor{reviewcolor}{RGB}{0,0,200}

\newcommand{\ceil}[1]{\lceil #1 \rceil}

\newcommand{\calL}{\mathcal{L}}
\newcommand{\calQ}{\mathcal{Q}}
\newcommand{\calP}{\mathcal{P}}

\DeclareMathOperator*{\argmin}{argmin}

\theoremstyle{plain}

\theoremstyle{definition}

\algdef{SE}[DOWHILE]{Do}{doWhile}{\algorithmicdo}[1]{\algorithmicwhile\ #1}%

\newcommand{\squishlist}{
 \begin{list}{$\bullet$}
  { \setlength{\itemsep}{0pt}
     \setlength{\parsep}{3pt}
     \setlength{\topsep}{3pt}
     \setlength{\partopsep}{0pt}
     \setlength{\leftmargin}{1.5em}
     \setlength{\labelwidth}{1em}
     \setlength{\labelsep}{0.5em} } }
     
\newcommand{\squishend}{
  \end{list}  }

\newcommand{\scatter}{\texttt{SCATTER}\xspace}

\graphicspath{{./figs/}}
\usepackage{geometry}
\copyrightyear{2022}
\acmYear{2022}
\setcopyright{acmcopyright}\acmConference[ICCAD '24]{Proceedings of the 43th IEEE/ACM International Conference of Computer-Aided Design (ICCAD)}{Oct 27 -- Oct 31, 2024}{Newark, NY, USA}
\acmBooktitle{Proceedings of the 59th ACM/IEEE Design Automation Conference (DAC) (DAC '22), July 10--14, 2022, San Francisco, CA, USA}
\acmPrice{15.00}
\acmDOI{10.1145/3489517.3530562}
\acmISBN{978-1-4503-9142-9/22/07}

\begin{document}

\expandafter\def\expandafter\normalsize\expandafter{%
    \normalsize%
    \setlength\abovedisplayskip{0pt}%
    \setlength\belowdisplayskip{8pt}%
    \setlength\abovedisplayshortskip{-8pt}%
    \setlength\belowdisplayshortskip{2pt}%
}

\settopmatter{printacmref=false} %

\pagestyle{plain} %

\title{
DarkLight: Dynamic Sparse Topology-Enabled Energy-Efficient, Robust Photonic Accelerator via in-situ Light Redistribution
}

\title{
L\textsuperscript{2}EGO: In-situ Light Refocusing and Gating Enabling Sparse Variable-Precision Photonic Tensor Cores
}

\title{
LightPATH: Sparse Reconfigurable Photonic Accelerators via Power-Efficient, Thermal-Aware In-situ Light Rerouting and Refocusing
}

\title{
ACCELight: \underline{A}lgorithm-\underline{C}ircuit \underline{C}o-Sparse Photonic Accelerator via Power-\underline{E}fficient, Thermal-Aware In-situ \underline{L}ight Refocusing
}

\title{
SCATTER: \underline{A}lgorithm-\underline{C}ircuit Co-\underline{S}parse Photonic Accelerator with \underline{T}hermal-\underline{T}olerant, Power-\underline{E}fficient In-situ Light \underline{R}edistribution
}

\author
{
Ziang Yin\textsuperscript{1}, Nicholas Gangi\textsuperscript{2}, Meng Zhang\textsuperscript{2}, Jeff Zhang\textsuperscript{1}, Rena Huang\textsuperscript{2},
Jiaqi Gu\textsuperscript{1}$^\dagger$\\
\textsuperscript{1}Arizona State University, \textsuperscript{2}Rensselaer Polytechnic Institute\\
\small\textit{$\dagger$jiaqigu@asu.edu}
}
\begin{abstract}
\label{abstract}
Photonic computing has emerged as a promising solution for accelerating computation-intensive artificial intelligence (AI) workloads. 
However, limited reconfigurability, high electrical-optical conversion cost, and thermal sensitivity limit the deployment of current optical analog computing engines to support power-restricted, performance-sensitive AI workloads at scale.
Sparsity provides a great opportunity for hardware-efficient AI accelerators. 
However, current dense photonic accelerators fail to fully exploit the power-saving potential of algorithmic sparsity.
It requires sparsity-aware hardware specialization with a fundamental re-design of photonic tensor core topology and cross-layer device-circuit-architecture-algorithm co-optimization aware of hardware non-ideality and power bottleneck.
To trim down the redundant power consumption while maximizing robustness to thermal variations, we propose \scatter, a novel algorithm-circuit co-sparse photonic accelerator featuring dynamically reconfigurable signal path via thermal-tolerant, power-efficient \emph{in-situ} light redistribution and power gating.
A power-optimized, crosstalk-aware dynamic sparse training framework is introduced to explore row-column structured sparsity and ensure marginal accuracy loss and maximum power efficiency.
The extensive evaluation shows that our cross-stacked optimized accelerator \scatter achieves a 511$\times$ area reduction and 12.4$\times$ power saving with superior crosstalk tolerance that enables unprecedented circuit layout compactness and on-chip power efficiency.
Our code is open sourced at \href{https://github.com/ScopeX-ASU/SCATTER}{link}.
\end{abstract}

\maketitle

\section{Introduction}
\label{sec:Introduction}
\begin{figure}
    \centering
    \includegraphics[width=\columnwidth]{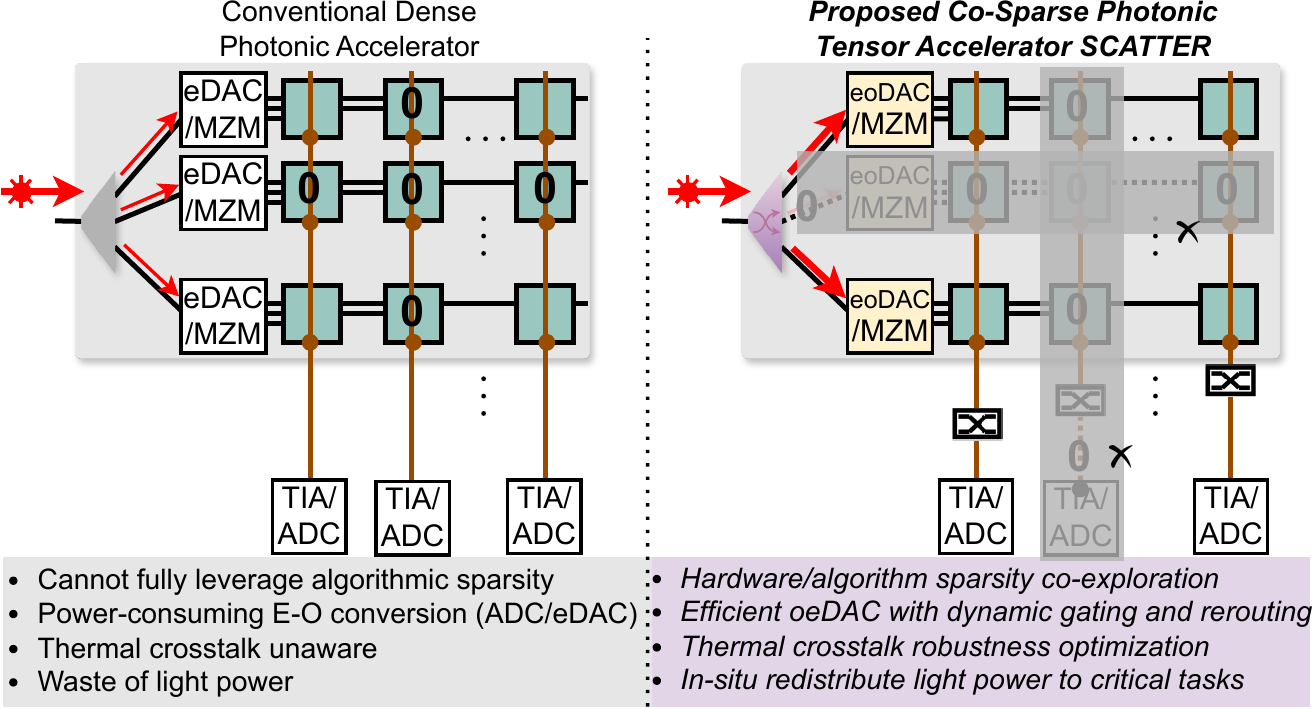}
    \vspace{-15pt}
    \caption{Our proposed \scatter architecture co-explores circuit/algorithm sparsity with power efficiency and robustness co-optimization compared to generic dense tensor cores.}
    \vspace{-10pt}
    \label{fig:Teaser}
    \vspace{-5pt}
\end{figure}

The quest for efficient and high-performance artificial intelligence (AI) solutions has propelled the development of photonic computing.
By harnessing the unique properties of light, photonic accelerators offer unmatched speed and energy efficiency, particularly for resource-constrained AI applications~\cite{NP_NATURE2017_Shen, NP_PIEEE2020_Cheng, NP_NaturePhotonics2021_Shastri, NP_ACS2022_Feng, NP_Science2024_Xu, NP_SciRep2017_Tait, NP_Nature2021_Xu, NP_Nature2021_Feldmann, NP_NatureComm2022_Zhu}.
Photonic tensor cores (PTCs) are the fundamental building blocks of these optical AI accelerators, and various designs have been demonstrated for matrix-vector multiplication or convolution using either coherent interference~\cite{NP_NATURE2017_Shen, NP_Science2024_Xu, NP_HPCA2024_Zhu, NP_ACS2022_Feng,NP_NatureComm2022_Zhu} or incoherent intensity modulation with multi-wavelength accumulation~\cite{NP_SciRep2017_Tait, NP_Nature2021_Xu, NP_Nature2021_Feldmann}.

Despite the ultra-fast speed and high throughput, the widespread adoption of photonic accelerators is hampered by several critical challenges: \emph{thermal robustness, non-trivial power bottleneck from signal conversion} between electrical and optical domains, and hardware reconfigurability ~\cite{NP_DATE2020_Gu, NP_ICCAD2019_Zhao, NP_ICCAD2020_Zhu,NP_TCAD2022_Mirza, NP_HPCA2024_Zhu,NP_arXiv2024_Zhang}. 
\ding{202}~\textbf{Thermal Variation Robustness}.~
Thermo-optic devices, chosen for their compactness and low insertion loss, are used for phase/magnitude modulation but are prone to thermal crosstalk, which significantly degrades computational accuracy.
Solutions often compromise chip density, require noise modeling~\cite{NP_DATE2020_Gu, NP_ICCAD2020_Zhu, NP_TCAD2022_Mirza}, or rely on on-chip calibration~\cite{NP_DAC2020_Gu, NP_NeurIPS2021_Gu,NP_JLT2024_Lu}, which induces hardware overhead and specific to individual chips.
A more general, thermal variation-tolerant architecture supporting standard model compression is needed.
\ding{203}~\textbf{E-O/A-D Conversion Power}.~
In electronic-photonic heterogeneous accelerators, signal conversion between optical and electrical domains is a significant power bottleneck~\cite{NP_NaturePhotonics2021_Shastri, NP_HPCA2024_Zhu,NP_DATE2020_Zokaee}. 
High-speed, high-resolution digital-to-analog converters (DACs) and analog-to-digital converters (ADCs) dominate on-chip power and area. Researchers have tried reducing power by lowering sampling frequency (<1 GHz) and bit resolutions (<4-bit) with minimal accuracy loss~\cite{NP_DATE2020_Gu, NP_arXiv2024_Zhang,NP_DATE2020_Zokaee}. 
However, achieving a balance between \emph{resolution, speed, and low power/area cost} remains challenging.
\ding{204}~\textbf{Reconfigurability}.~
Reprogrammability is crucial for versatile photonic accelerators. 
Current PTC designs are specialized for dense MVM with fixed circuit topologies, lacking flexibility to exploit sparsity in modern AI models. 
An efficient co-sparse architecture should optimize more than just setting weights to zero. 
A reconfigurable PTC that adaptively reroutes and gates its signal path to support algorithmic sparsity would significantly enhance flexibility and efficiency.

To address these fundamental roadblocks, \emph{for the first time}, we present a dynamically reconfigurable photonic accelerator \scatter that features native support for algorithm/hardware co-sparsity with cross-layer power/thermal robustness co-optimization.
\ding{202} \underline{To boost the} \underline{thermal variation robustness}, we optimize device spacing, exploit circuit sparsity, and employ \emph{in-situ} power gating to minimize crosstalk.
\ding{203} \underline{To boost power efficiency}, we
explore power-optimized photonic devices, hybrid electrical-optical DAC designs, and architectural hardware sharing.
Our power-aware dynamic sparse training framework explicitly targets power minimization during sparsity exploration.
\ding{204} \underline{To enable flexible hardware reconfiguration}, we introduce an on-chip tunable light rerouter to dynamically redistribute the optical power to efficiently support structured row-column weight sparsity that directly translates to power reduction and noise suppression.

The major contributions of this paper are as follows:

\squishlist
     {\item \textbf{\emph{In-situ} Light Redistribution} -- \emph{For the first time}, we introduce \emph{in-situ} light redistribution mechanism for reconfigurable photonic tensor cores, achieving power-optimized, thermal-robust algorithm-circuit co-sparsity. We dynamically reroute light power to focus on critical computations and suppress variation-induced errors from pruned components, enhancing efficiency and signal-to-noise ratio.}
    {\item \textbf{Dynamic Reconfigurable Architecture} -- We introduce on-chip optical rerouter input/output power gating to enable fine-grained signal path control as a native primitive for circuit sparsity, achieving superior efficiency for multi-core photonic AI accelerators.}
    {\item \textbf{Cross-layer Power/Area Minimization} -- We integrate power-optimized photonic devices, hybrid electrical-optical DAC designs, dynamic power-gated input/readout circuitry, and automated architecture exploration, realizing 511$\times$ area reduction and 12.4$\times$ power saving compared to dense designs built on foundry devices.}
    {\item \textbf{Power-Robustness Co-Optimized Sparsity} -- Our one-shot, hardware-aware dynamic sparse training learns structured weight sparsity masks while optimizing accuracy, power efficiency, and thermal crosstalk tolerance.}
\squishend

\vspace{-10pt}

\section{Background}
\label{sec:Background}
\subsection{Dense/Sparse Optical Neural Networks}
\label{sec:PhotonicsBasics}
Various photonic neural network designs encode inputs and weights to light magnitude/phase and circuit transmission, performing ultra-fast matrix multiplication~\cite{NP_NATURE2017_Shen, NP_PIEEE2020_Cheng, NP_NaturePhotonics2021_Shastri, NP_ACS2022_Feng,NP_SciRep2017_Tait, NP_Nature2021_Xu, NP_Nature2021_Feldmann, NP_NatureComm2022_Zhu,NP_Science2024_Xu, NP_HPCA2024_Zhu, NP_ACS2022_Feng}. 
However, most prior work focuses on dense photonic tensor cores (PTCs) with fixed topologies, limiting compatibility with the algorithmic sparsity in modern AI models. 
Some pruning techniques for optical neural networks have shown power reduction by pruning phase shifters in MZI arrays~\cite{NP_ASPDAC2020_Gu, NP_TCAD2022_Gu, NP_ISVLSI2022_Banerjee}, but they fail to fully leverage structured sparsity. 
The challenge remains to dynamically reconfigure circuit connectivity and signal paths for co-exploration of algorithmic and circuit sparsity, optimizing power and robustness.

\subsection{Structured Sparsity and Dynamic Sparse Training}
Modern neural networks often exhibit intrinsic sparsity, offering opportunities for memory and computational savings through pruning~\cite{NN_NIPS2016_Wen,NN_HPCA2020_Zhang,NN_PIEEE2020_Deng,NN_JSSC2021_Zhang}.
Unlike unstructured sparsity with arbitrary zero entries in the matrix, structured sparsity, where clusters of elements are pruned in hardware-aware patterns, is particularly advantageous for efficient implementation~\cite{NN_NIPS2016_Wen,NN_PIEEE2020_Deng,NN_arXiv2020_Lym}.

To automate structured sparsity exploration while minimizing accuracy loss, we adopt state-of-the-art dynamic sparse training (DST)~\cite{NN_NeurIPS2021_Chen,NN_ICLR2024_Lasby,NN_ICLR2020_Liu_DST}. 
Unlike traditional approaches that first train a dense model, DST maintains a sparse model throughout training, iteratively pruning and regrowing connections. This one-shot approach streamlines the neural architecture search process.
We will develop our power/robustness sparsity optimization based on the flexible DST framework.

\section{Proposed Co-Sparse Photonic Accelerator \scatter }
\label{sec:Method}

\begin{figure}
    \centering
    \includegraphics[width=\columnwidth]{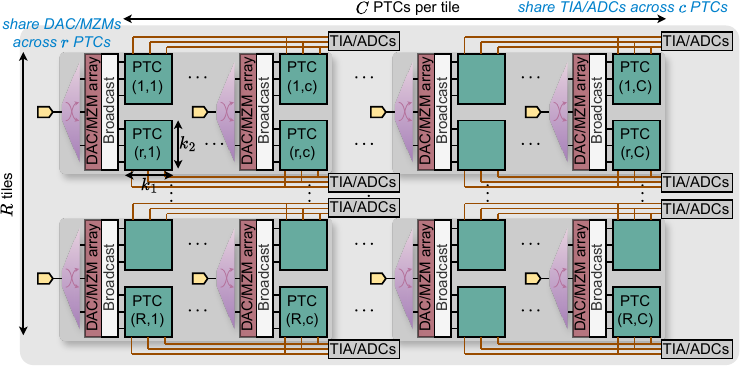}
    \vspace{-10pt}
    \caption{Dynamic multi-core photonic accelerator architecture with $R$ tiles and $C$ PTCs per tile. Each PTC is of size $k_1 \times k_2$.
    Input modulation modules are shared by $r$ PTCs across different tiles.
    Readout circuitry is shared by $c$ PTCs in a tile.}
    \label{fig:ArchOverview}
\end{figure}

We introduce \scatter, a multi-core dynamic photonic accelerator, shown in Fig.~\ref{fig:ArchOverview}.
\scatter is designed to overcome the limitations of traditional photonic accelerators with these key features: 
\ding{202} phase-agnostic incoherent photonic tensor cores for robust tensor computing; 
\ding{203} shared input modulation modules and readout circuitry to balance area, power, and control flexibility; 
\ding{204} in-situ tunable rerouter for light redistribution; 
\ding{205} hardware gating to support structured row-column sparsity with enhanced power efficiency and thermal crosstalk robustness.; 
\ding{206} co-optimized devices, circuits, and architecture configurations with maximum efficiency and thermal variation tolerance.
In this section, we will detail the core innovations of \scatter's hardware/algorithm co-sparse design and our comprehensive cross-layer co-optimization methodology.

\subsection{Accelerator Architecture Overview}
\label{sec:ArchOverview}

\subsubsection{Phase-Insensitive Differential Photonic Tensor Cores}
\label{sec:TensorCore}
\begin{figure}
    \centering
    \includegraphics[width=0.82\columnwidth]{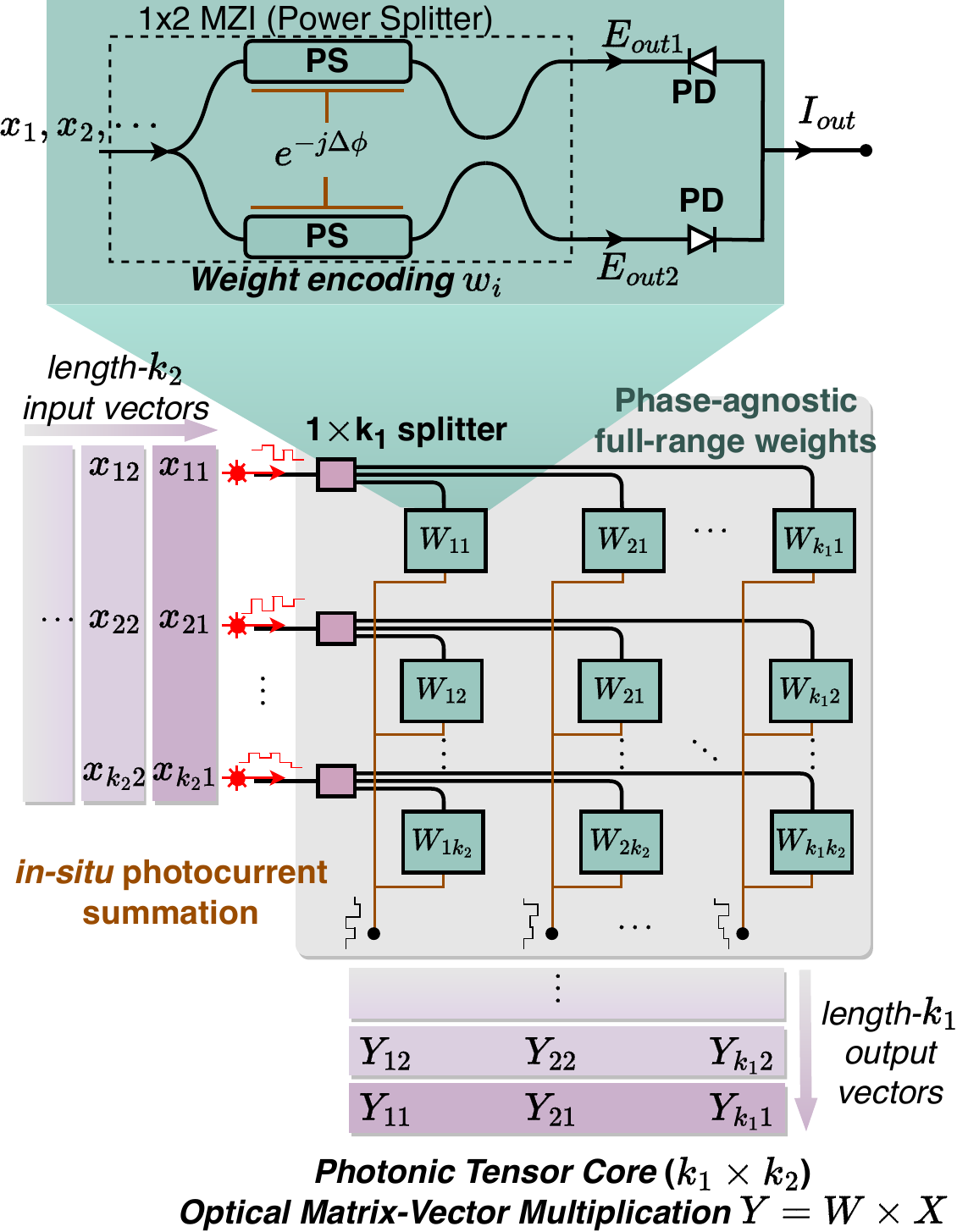}
    \vspace{-10pt}
    \caption{Schematic of phase-agnostic incoherent PTC.}
    \label{fig:TensorCore}
    \vspace{-10pt}
\end{figure}

To avoid the phase instability issue of coherent PTCs, e.g., unitary MZI mesh~\cite{NP_NATURE2017_Shen}, and dynamic crossbar arrays~\cite{NP_HPCA2024_Zhu}, and thermal sensitivity issue of narrow-band resonance-based PTCs, e.g., MRR weight banks~\cite{NP_SciRep2017_Tait}, we introduce a phase-agnostic full-range PTC architecture.
This architecture forms the foundation of our sparsity optimization techniques.
Figure~\ref{fig:TensorCore} illustrates a $k_1 \times k_2$ PTC.
The length-$k_2$ input vector $x$ is encoded as light intensity and broadcast to $k_1$ columns via 1$\times k_1$ even splitter.
Each crossbar node is a full-range multiplication engine consisting of a 1$\times$2 MZI power splitter and balanced photodetectors (BPD).
Partial products are accumulated along each column through photocurrent aggregation.
Formally, the tensor core operation is
\begin{equation}
    \small
    \label{eq:TensorCore}
    \begin{aligned}
    y=Wx; y_i&=\sum_jW_{ij}x_{j};\\
    \begin{pmatrix}
        E_{out1}\\
        E_{out2}
    \end{pmatrix}&=
    \frac{\sqrt{2}}{2}\begin{pmatrix}
        1 & j\\
        j & 1
    \end{pmatrix}
    \begin{pmatrix}
        e^{-j\Delta\phi} & 0\\
        0 & 1
    \end{pmatrix}
    \begin{pmatrix}
        \sqrt{2}/2\\
        \sqrt{2}/2
    \end{pmatrix}x\\
    W_{ij}x=|E_{out1}|^2-|E_{out2}|^2&=\Big(2\cos^2{\big(\frac{\Delta\phi+\phi_b}{2}\big)-1}\Big)x,    
    \end{aligned} 
\end{equation}
where $\Delta\phi\in[-\frac{\pi}{2},\frac{\pi}{2}]$ is the MZI arm phase difference.
The default phase bias $\phi_b$ is $\frac{\pi}{2}$.
Differential outputs from the power splitter and BPDs enable full-range weight representation.
For input vectors, since they are quantized within a certain range, we can adopt non-negative isomorphic transformation before deployment to guarantee positive-only input $x$ with a certain bias~\cite{NP_Arxiv2023_Kirtas}.
The weight matrices and input vectors are normalized to ensure they are implementable by the modulation coefficient and light intensity, and the output results are scaled back with the normalization factor.
Note that neither phase coherence nor thermal feedback, like in MZI and MRR arrays, is needed due to intensity encoding and broadband non-resonance devices.

\subsection{Power, Area, and Robustness Analysis}
\label{sec:PowerRobustAnalysis}
We present a thorough power analysis of photonic computing engines, which will guide our optimization strategies.
We assume the multi-core architecture has $R$ tiles and $C$ cores per tile, operating at frequency $f$. 
The input modulation module is shared across $r$ PTCs across different tiles, and the readout circuitry is shared across $c$ cores within a tile.

\subsubsection{On-chip Power Modeling}
\label{sec:PowerModeling}
We break down the key contributors to on-chip power consumption:

For input modulation, high-speed $b_{in}$-bit DACs consume a large amount of power. 
The input modulation power is estimated as
\begin{equation}
    \small
    \label{eq:InputPower}
    \begin{aligned}
    P_{in}&=\frac{RCk_2}{r}(P_{mod}+P_{eDAC}(b_{in},f)),\\
    P_{mod}&=P_{mod,static}+E_{mod}f,~~
    P_{eDAC}(b_{in},f)=P_{0,eDAC}\frac{2^{b_{in}}}{b_{in}+1}f,
    \end{aligned}
\end{equation}
where $P_{mod,static}$ is the static power of the MZM, $E_{mod}$ is the dynamic energy per full-range modulation (J/bit), $P_{0,eDAC}$ is the reported eDAC power working at its designed sampling rate and precision.
Note that the eDAC power scales linearly with frequency $f$ and exponentially with resolution $b_{in}$.
Reducing the eDAC power is crucial for system energy efficiency.

For weight encoding, phase shifters in MZI splitters, low-speed ($f_w \ll f$) $b_w$-bit weight DACs, and BPDs contribute to the weight encoding power:
\begin{equation}
    \small
    \label{eq:WeightPower}
    \begin{aligned}
    P_{wgt}&=RCk_1k_2\big(P_{MZI} + 2P_{PD}\big), ~ P^{ij}_{PS}=\calP( |\Delta\phi|,l_s),
    \end{aligned}
\end{equation}
where the MZI static power dominates.
This power is a function of the absolute phase difference $|\Delta\phi|=|\Delta\phi^{up}-\Delta\phi^{lo}|$ and MZI arm spacing $l_s$.
The simulated power function $\calP(\cdot)$ is shown in Fig.~\ref{fig:CrosstalkMZIPower}.

The readout circuitry also consumes significant power, especially for the high-speed ADCs.
The total readout power is,
\begin{equation}
    \small
    \label{eq:WeightPower}
    \begin{aligned}
    P_{out}&=\frac{RCk_1}{c}\big(P_{TIA}+P_{ADC}(b_{o},f)\big), P_{ADC}(b_o,f)=P_{0,ADC}\cdot b_of,
    \end{aligned}
\end{equation}
where the ADC power dominates the readout power, which scales linearly with output precision and sampling frequency.
Hence, the total on-chip power is $P=P_{in}+P_{wgt}+P_{out}$.
Note that off-chip laser and low-speed weight DACs are not included in this model.

\subsubsection{Area Modeling}
\begin{figure*}
    \centering
    \includegraphics[width=\textwidth]{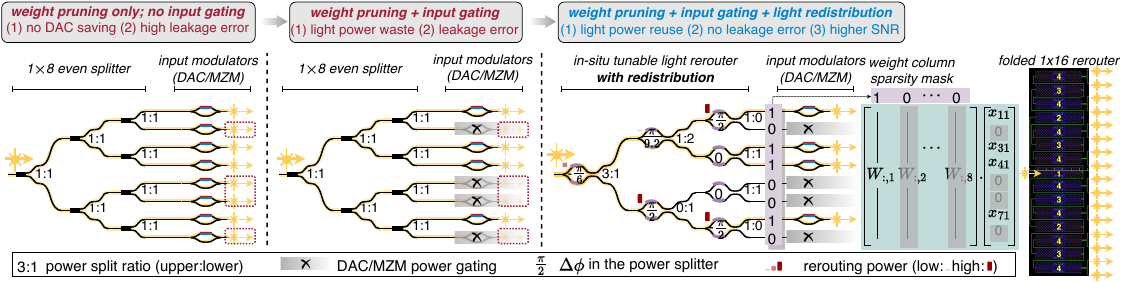}
    \vspace{-12pt}
    \caption{Weight block column-wise sparsity can be supported by on-chip light rerouter with in-situ tunable light splitting ratios. 
    Here, we show an 8$\times$8 block as an example.
    Input gating helps save significant high-speed DAC and input modulation power while reducing leakage error in pruned paths.
    Light redistribution eliminates leakage errors and provides light power to unpruned computing engines with higher optical SNR.
    Different from the tree structure in the schematic, a folded rerouter layout is designed to save area.
    Refocusing can effectively reduce computing N-MAE errors compared to standard weight pruning.}
    \label{fig:TunableRerouter}
    \vspace{-5pt}
\end{figure*}

Each crossbar node area is
\begin{equation}
    \small
    \label{eq:NodeArea}
    A_{node}=(l_s+w_{PS})\times (l_{Y}+l_{PS}+l_{DC}),
\end{equation}
based on our designed phase shifter and layout, $(l_{Y}+l_{PS}+l_{DC})=115 \mu m$ and $w_{PS}=6 \mu m$.
The $k_1 \times k_2$ array area is
\begin{equation}
    \small
    \label{eq:NodeArea}
    A_{PTC,wgt}=\big((k_2-1)l_v+l_{Y}+l_{PS}+l_{DC}\big)\times \big((k_1-1)l_h+ls+w_{PS}\big).
\end{equation}
Considering the multi-core architecture with an input modulation sharing factor $r$ and readout sharing factor $c$, the total on-chip area is estimated as
\begin{equation}
    \small
    \label{eq:ArchArea}
    \begin{aligned}
        &A=RC(A_{PTC,wgt}+k_2A_{MMI}+2k_1k_2A_{PD})\\
        &+\frac{RC}{r}(k_2A_{DAC}+k_2A_{MZM}+A_{rerouter})
        +\frac{RC}{c}(k_1A_{ADC}+k_1A_{TIA}),
    \end{aligned}
\end{equation}
where the PD arrays are placed in separate regions to avoid thermal noises, $A_{MMI}$ is the 1$\times k_1$ MMI splitter, $A_{rerouter}$ is the area corresponding to the compact folded layout shown in Fig.~\ref{fig:TunableRerouter}.
Note that laser and weight DACs are off-chip and are not included here.

\subsubsection{Thermal Crosstalk Modeling}

\begin{figure}
    \centering
    \vspace{-10pt}
    \subfloat[]{\includegraphics[width=0.49\columnwidth]{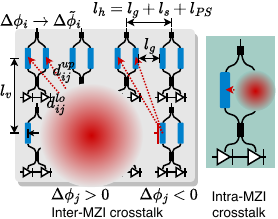}
    \label{fig:Crosstalk}
    }
    \subfloat[]{\includegraphics[width=0.31\columnwidth]{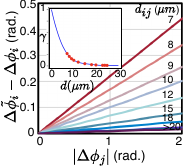}
    \label{fig:CrosstalkGammaCurve}
    }
    \subfloat[]{\includegraphics[width=0.2\columnwidth]{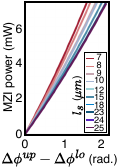}
    \label{fig:CrosstalkMZIPower}
    }\\
    \vspace{-10pt}
    \subfloat[]{\includegraphics[width=0.29\columnwidth]{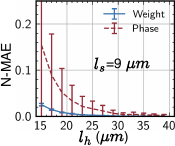}
    \label{fig:CrosstalkWeightError}}
    \subfloat[]{\includegraphics[width=0.69\columnwidth]{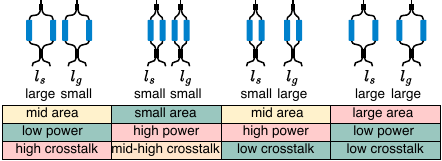}
    \label{fig:CrosstalkSpacingCompare}}
    \vspace{-10pt}
    \caption{(a) Inter- and intra-MZI thermal crosstalk are modeled by distance-related coupling coefficients $\gamma$.
    (b) Lumerical HEAT simulation is used to sweep various phase shifter spacings and fit a numerical crosstalk model.
    (c) Larger arm spacing $l_s$ reduces the required MZI power to realize the same phase difference.
    (d) Larger MZI spacing $l_h$ reduces normalized mean-absolute error (N-MAE) on phases and weights.
    (e) Impact of arm spacing and MZI spacing on area, power, and crosstalk.
    }
    \label{fig:CrosstalkAnalysis}
    \vspace{-10pt}
\end{figure}

Thermo-optic MZI power splitters experience intra-MZI and inter-MZI thermal crosstalk, which leads to power penalty and accuracy degradation. 
In a $k_2$-row and $k_1$-column PTC, we model the impact from other MZIs to the $i$-th MZI as follows,
\begin{equation}
    \small
    \label{eq:Crosstalk}
    \begin{aligned}
        \Delta\widetilde{\phi}_i=\Delta\phi_i+\sum_{j\neq i}^{k_1k_2}\Delta\gamma_{ij}|\Delta\phi_j|=\Delta\phi_i+\sum_{j\neq i}^{k_1k_2}\big(\gamma_{ij}(d_{ij}^{up})-\gamma_{ij}(d_{ij}^{lo})\big)|\Delta\phi_j|,
    \end{aligned}
\end{equation}
where $\Delta\phi_i$ is the target phase shift, and $\Delta\gamma_{ij}$ is the crosstalk coefficient between the $i$-th and $j$-th MZIs.
This coefficient accounts for the differential working mode and depends on the distance between the aggressor and victim phase shifters.
$\gamma_{ij}$ is a function of center distance $d_{ij}$ between the aggressor phase shifter in the $j$-th MZI and the victim phase shifter in the $i$-th MZI.
$d_{ij}^{up}$ and $d_{ij}^{lo}$ represent the distance w.r.t the upper and lower arm of the victim MZI, respectively.

This distance is calculated dynamically based on the sign of the aggressor's phase shift $\Delta\phi_j$.
It will heat up the upper arm to realize a positive $\Delta\phi\in[0,\pi/2]$ and heat up the lower arm to create a negative $\Delta\phi\in[0,-\pi/2]$.
We formulate the phase-dependent distance as
\begin{equation}
    \small
    \label{eq:CrosstalkDistance}
    \begin{aligned}
        d_{ij}^{up}=\sqrt{[(R(j)-R(i))l_v]^2+[(C(j)-C(i))h-l_s\mathbf{I}_{\Delta\phi_j<0}]^2},\\
        d_{ij}^{lo}=\sqrt{[(R(j)-R(i))l_v]^2+[(C(j)-C(i))h+l_s\mathbf{I}_{\Delta\phi_j\ge 0}]^2},
    \end{aligned}
\end{equation}
where the indicator function $\textbf{I}_{\Delta\phi_j<0}$ is 1 with negative $\Delta\phi_j$, and $C(\cdot)$ and $R(\cdot)$ are the column and row index of $j$-th MZI.

We use Lumerical HEAT and MODE simulations to characterize the relationship between $\gamma$ and $d$, shown in Fig.~\ref{fig:Crosstalk}.
Thermal profiles were imported into MODE to determine the effective indices and $\Delta\phi$ of the upper and lower arms based on the thermo-optic coefficient of silicon.  
With the same spacing, the crosstalk factor $\gamma\propto\frac{\Delta\phi_i}{\Delta\phi_j}$ is constant, which indicates that $\gamma$ is only a function of spacing.
The crosstalk coefficient decays exponentially with increasing distance.
We fit this relationship with a piecewise function (5th-order polynomial and exponential function),
\begin{equation}
    \small
    \label{eq:CrosstalkCurve}
    \gamma(d)=(\sum_{i=0}^5p_id^i)\mathbf{I}_{d<23} + a_0e^{-a_1d}\mathbf{I}_{d\ge 23},
\end{equation}
where the coefficients are [$p_0, \cdots, p_5$]=[1,-1.76e-1,9.9e-3,-8.30e-6,-1.56e-5,3.55e-7], [$a_o, a_1$]=[0.217, 0.127]. The curve fitting fidelity $R^2$ are 0.999 for the polynomial part and 0.998 for the exponential part.

\subsection{Power, Area, and Robustness Co-Optimization}
We introduce synergistic optimization approaches across device, circuit, architecture, and algorithm levels, guided by our in-depth efficiency and robustness analysis.
\subsubsection{Device-Level: Power-Efficient Footprint-Compact MZI}
Foundry-provided MZI switch (Foundry-MZI) consumes $P_{\pi}$=30 mW for $\pi$ phase shift and a large footprint consumption (550 $\mu m$ in length).
We design a low-power MZI switch (LP-MZI) with compact size (115 $\mu m$ in length) and 50\% lower power $P_{\pi}$=15 mW.
The phase shifter width $w_{PS}$ and phase shifter spacing $l_s$ between two arms impact the switching power and device footprint.
The phase difference between the upper and lower arm of an MZI is $\Delta\phi^{up} -\Delta\widetilde{\phi}^{lo}$.
If the upper arm is heated up, the intra-MZI crosstalk will increase $\Delta\widetilde{\phi}^{lo}$ and thus diminish the phase difference.
It results in a power penalty required to realize the same $\Delta\phi$.
We show how MZI power $P_{MZI}$ changes with arm spacing and phase difference between arms in Fig.~\ref{fig:CrosstalkMZIPower}.
Later, we will explore optimal settings to balance power and area and show our efficiency advantage over standard foundry devices.

\begin{figure}
    \centering
    \subfloat[]{\includegraphics[width=0.52\columnwidth]{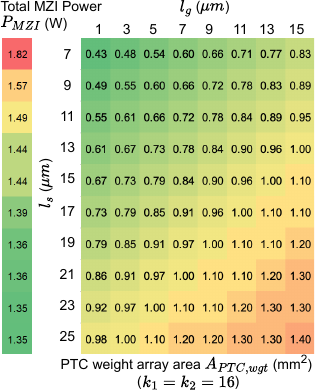}
    \label{fig:CrosstalkPowerArea}
    }
    \subfloat[]{\includegraphics[width=0.45\columnwidth]{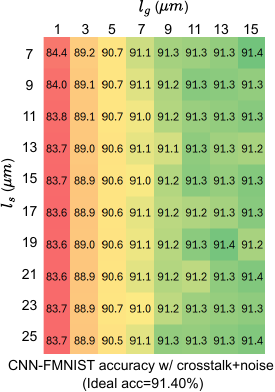}
    \label{fig:CrosstalkAccuracy}
    }
    \vspace{-5pt}
    \caption{(a) Area and power of weight MZI array ($k_1=k_2=16$) with different arm spacing $l_s$ and MZI gap $l_g$.
    (b) Accuracy under variations on CNN FashionMNIST with different spacing.}
    \label{fig:CrosstalkPowerAreaAcc}
\end{figure}

Figure~\ref{fig:CrosstalkPowerAreaAcc} illustrates the power-area-accuracy design space of a $16 \times 16$ PTC.
We carefully select device spacing configurations ($l_s$, $l_g$) to balance power, area, and accuracy.
For a dense PTC, to meet the accuracy target (e.g., <1\% drop), we set $l_s=9\: \mu m$, $l_g=5\: \mu m$ by minimizing the power-area product.
Importantly, we'll demonstrate how sparsity techniques can relax these design constraints and allow for an even more compact layout, leading to improvements in both power and area efficiency.

\subsubsection{Circuit-Level: In-situ Tunable Light Redistribution for Column Sparsity}
\label{sec:ReRouter}
High-speed DACs and drivers for input modulation consume significant power. 
To maximize efficiency, we strategically zero out length-$rk_1$ column vectors in the $rk_1 \times ck_2$ weight chunk, enabling us to shut down the weight MZIs on those pruned rows.
However, due to non-idealities (e.g., phase bias deviation, crosstalk, and noises), simply removing power from pruned weight MZIs can still lead to non-zero weights, which induces computing errors.

We propose a dynamically reconfigurable sparse tensor core with in-situ light redistribution. This is the key to fully leveraging sparsity benefits.
To illustrate this, we first express the vector dot-product result with crosstalk and noise as:
\begin{equation}
    \small
    \label{eq:DotProduct}
    y=\sum_{j}^{k_2} (\widetilde{w}_jx_j + \delta n_{PD}),
\end{equation}
where $\delta n_{PD}$ is random photocurrent noises from PDs (we set it to 0.01), and $\widetilde{w}$ is the weight under crosstalk.
Given a weight column sparsity mask $m^c=[m^c_1,\cdots,m^c_{k_2}]\in{\{0,1\}}^{k_2}$, 
we assume there are $k_2'$ nonzero elements in $m^c$.

Based on the above assumption, we compare two conventional design approaches and highlight our superiority with \emph{in-situ} light redistribution technique in Fig.~\ref{fig:TunableRerouter}.

\noindent\textbf{\underline{Weight Pruning Only}}.~
In Fig.~\ref{fig:TunableRerouter}(\emph{Left}), the input light is evenly split via a balanced splitter tree without shutting down modulators.
Though weight MZI power is removed, the DAC/MZM power is wasted.
More importantly, pruned paths still contribute to the final photocurrent, leading to leakage errors:
\begin{equation}
    \small
    \label{eq:DotProductWeightPrune}
    y_{\text{prune}}=\sum_{j,m^c_j=1}^{k_2} (\widetilde{w}_jx_j)+\sum_{j,m^c_j=0}^{k_2} (\delta w_j\cdot x_j)+\sum_{j}^{k_2} \delta n_{PD},
\end{equation}
where $\delta w$ is the error due to non-idealities, e.g., weight MZI crosstalk, random phase noises, and limited extinction ratio (ER) of MZIs, defined as the ratio between maximum and minimum transmission.

\noindent\textbf{\underline{Weight Pruning + Input Gating (IG)}}.~
In Fig.~\ref{fig:TunableRerouter}(\emph{Middle}), the power supply of the high-speed DACs and MZMs for pruned ports are gated. 
While this saves some power, light still leaks through the high-speed MZMs (due to a limited extinction ratio). 
Further, light power on pruned paths is completely wasted without contributing to useful computation. The dot-product result is as follows
\begin{equation}
    \small
    \label{eq:DotProductPruneGating}
    y_{\text{IG}}=\sum_{j,m^c_j=1}^{k_2} (\widetilde{w}_jx_j)+\sum_{j,m^c_j=0}^{k_2} (\delta w_j\cdot \delta x_j)+\sum_{j}^{k_2} \delta n_{PD},
\end{equation}
where $\delta w$ and $\delta x$ are nonzero errors due to non-ideal variations.
Note that column pruning with input MZM gating has no reduction in the PD noises and still suffers from leakage errors, as in the second term.
\noindent\textbf{\underline{Pruning + Input Gating + Light Redistribution (LR)}}.
As shown in Fig.~\ref{fig:TunableRerouter}(\emph{Right}), our proposed solution upgrades the passive even splitter tree to an \emph{in-situ} tunable light rerouter to dynamically redistribute light power from unused ports to actives ports.

This boosts the intensity on active ports by a factor of $k_2/k_2'$. 
The TIA gain will be reduced by a ratio of $k_2'/k_2$ to recover the same range.
Then, the result becomes
\begin{equation}
    \small
    \label{eq:DotProductRefocus}
    y_{\text{IG+LR}}=\frac{k_2'}{k_2}\Big(\!\!\!\sum_{j,m^c_j=1}^{k_2} (\frac{k_2}{k_2'}\widetilde{w}_jx_j)+\sum_{j}^{k_2} \delta n_{PD}\Big)=\!\!\!\sum_{j,m^c_j=1}^{k_2}\!\!\!\widetilde{w}_jx_j+\frac{k_2'}{k_2}\sum_{j}^{k_2} \delta n_{PD}.
\end{equation}
Light redistribution has two main advantages: it eliminates the leakage errors on pruned ports and effectively reduces the photocurrent noise by a factor of $\frac{k_2'}{k_2}$.
For example, with a 20\% column sparsity, light redistribution will have a 7 dB higher SNR.
Optical power redistribution through light path reconfiguration can be realized by cascading a number of MZI power splitters.
Later, our power-efficient sparse training algorithm will find the optimal column sparsity mask $m^c$ that minimizes the power consumption of the rerouter given a defined column sparsity.

\begin{figure}
    \centering
    \includegraphics[width=0.95\columnwidth]{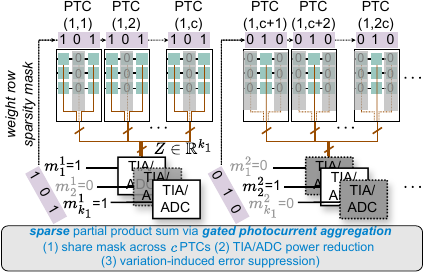}
    \vspace{-5pt}
    \caption{Weight block matrix row-wise sparsity with output TIA/ADC gating (OG).
    }
    \label{fig:Switches}
    \vspace{-5pt}
\end{figure}

\subsubsection{Circuit-Level: On-chip TIA/ADC Gating for Row Sparsity}
\label{sec:Switches}

To maximize power savings from sparsity, we enable dynamic TIA/ADC gating in Fig.~\ref{fig:Switches}.
Our architecture accumulates the photocurrent from $c$ PTCs per tile as analog-domain partial product summation and shares the same TIA/ADC array among them.
To fully exploit the benefits of power gating, we focus on coarse-grained structured sparsity where entire length-$ck_2$ row vectors in the $rk_1\times ck_2$ weight chunk are pruned.
This allows us to \emph{shut down corresponding TIA/ADC for energy reduction and eliminate any leakage, crosstalk, and PD noises}.

\subsubsection{Circuit-Level: Efficient Hybrid Electronic-Optic DAC}
\begin{figure*}
    \centering
    \includegraphics[width=\textwidth]{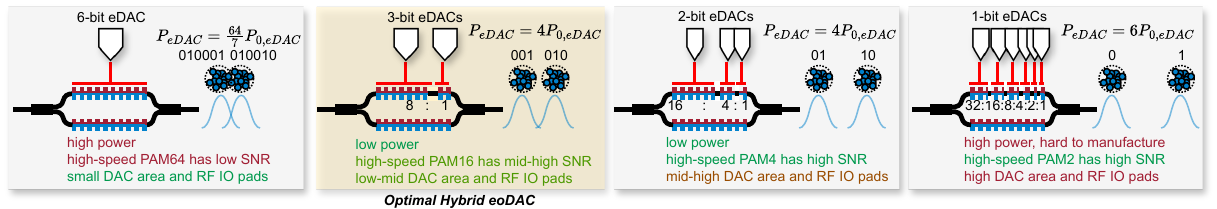}
    \vspace{-15pt}
    \caption{Hybrid electrical-optical DAC (eoDAC) with optimal settings gives the best power, area, manufacturability, and SNR.}
    \label{fig:oDAC}
\end{figure*}

\label{sec:oDAC}
High-speed DACs are a major power bottleneck and often limit resolution and signal-to-noise ratio (SNR).
As shown in Fig.~\ref{fig:oDAC}, generating 6-bit PAM signals at 5 GHz with a single electronic DAC (eDAC)  incurs a high power cost $\frac{64}{7}P_{0,eDAC}$ and suffers from low SNR due to overlapped symbols.
We introduce a hybrid electronic-optic DAC (eoDAC) that realizes weighted signal modulation both in the electrical and optical domains.
The modulation actuators are partitioned into nonuniform segments, and each can be modulated with a low-bit eDAC.
There exist fundamental trade-offs among eDAC area, number of IO pads, eDAC power, SNR, and manufacturability.
Figure~\ref{fig:oDAC} shows different hybrid eoDAC settings.
We find that an optimal design partitions the phase shifter into two segments (with an 8:1 length ratio) controlled by 3-bit eDACs.
It can approximately realize a 6-bit PAM signal via two 3-bit modulators, e.g., $010001=2^3\cdot(010)+001$.
The length ratio can be customized based on the actual MZM response.
This setting requires twice the independent IO pads but saves 2.3$\times$ DAC power with significant SNR improvement.
Further partitioning (e.g., pure optical DAC) offers negligible power benefits while increasing area, layout, and manufacturing complexity.

\subsubsection{Algorithm-Level: Power/Crosstalk-Aware Dynamic Sparse Training}
\label{sec:DSTAlgo}
We adapt the SoTA DST algorithm to automatically select the structured sparsity patterns aware of accuracy, power, and crosstalk.
Unlike conventional pruning methods, which start from a dense pre-trained model, we initialize a sparse model and dynamically explore sparsity patterns, balancing accuracy, power, and robustness.

\noindent\textbf{\underline{Crosstalk/Power-Minimized Initialization}}.~
Assume a model has $L$ convolutional (CONV) layers, and sparsity is not applied to the first CONV layer and the last linear layer.
For the $l$-th CONV layer, the weight matrix is of size $W^l\in\mathbb{R}^{C_{o}C_{i}KK}$.
After \emph{im2col}, the unfolded weight matrix will be padded and partitioned into a 6-D tensor $W^l\in\mathbb{R}^{p\times q\times r\times c \times k_1 \times k_2}$, where $p=\ceil{\frac{C_o}{rk_1}}$ and $q=\ceil{\frac{C_iK^2}{ck_2}}$.
With a given sparsity $s$ (percentage of nonzero elements), we need to first assign the layer-wise sparsity $(s^1,\cdots, s^L)$ to match the target sparsity and then initialize the sparsity mask for the $l$-th layer $m^l=(m^{(l,c)},m^{(l,r)})$, which contains a column mask $m^c\in\{0,1\}^{p\times q \times r \times 1 \times k_1 \times 1}$ and a row mask $m^c=r\in\{0,1\}^{p\times q \times 1\times c \times 1 \times k_2}$.

For simplicity, we assign the same sparsity for all layers.
Better strategies in the literature can be applied.
With a target sparsity $s_l$, we assume the same sparsity pattern for each $k_1\times k_2$ weight block.

Since the horizontal spacing $l_h<20\;\mu m$ is much smaller than the vertical spacing $l_v=120\;\mu m$, crosstalk primarily occurs between weights in the same column vector.
We initialize the row mask $m^r$ with an interleaved pattern (e.g., 101010...) to minimize crosstalk based on the guidance from Fig.~\ref{fig:RowSparsityOG}(a).
The row mask requires up to 50\% row sparsity, which contains the maximum 0's needed to eliminate crosstalk.
If the target sparsity is high ($s>0.5$), we allocate all sparsity to the row mask for maximum crosstalk reduction.
E.g., $s^r=0.75$ and $rk_1=8$ lead to a row mask of 11111010.
If $s<0.5$, we initialize column mask $m^c$ with a column sparsity of $s^c=s/s^r$ followed by power optimization.
For each $rk_1 \times ck_2$ block, among $ck_2$ column vectors, we find the best group of $[ck_2s^c]$ vectors among all $\binom{ck_2}{[ck_2s^c]}$ combinations with the lowest power.

\emph{\underline{How to Calculate Power Metric for a Mask?}}~
Power $P$ is estimated based on Section~\ref{sec:PowerModeling}.
In \scatter, the sparsity mask indicates power gating on corresponding weight MZIs, input modulation modules, and readout TIA/ADC arrays to save power and reduce leakage noises.
The power of the light rerouter can be calculated by the power splitting ratio derived from the column mask.
E.g., with a mask $m^c=10110010$, the root splitter's ratio is \emph{up:lo}=(1+0+1+1)/(0+0+1+0)=3:1, and the phase is $\Delta\phi=2\text{arccos}(\sqrt{\frac{up}{lo+up}})-\phi_b$.
If \emph{up+lo}=0, we set $\Delta\phi=0$.
Then, its power can be obtained by using the $\calP(|\Delta\phi|, l_s)$ function.
Similarly, the weight MZIs' power is dynamically calculated with the actual weights and defined arm spacing.

\noindent\textbf{\underline{Power-Aware Pruning Procedure}}.~
For simplicity, we only discuss one layer and remove the layer index $l$ in all notations.
As shown in Alg.~\ref{alg:DSTAlgo}, for every $\Delta T$ step, if $s^c<1$, we fix the row mask and explore column sparsity pattern with one pruning and one growth stage, 
We define a death rate $\alpha$ that controls the percentage of unpruned weights to be pruned in the current step.
We apply step-wise CosineDecay to death rate $\alpha^{t+1} \gets \frac{\alpha^{t}}{2}(1+\cos(\frac{t\pi}{T_{end}}))$ for stable exploration and convergence, where $T_{end}$ is 80\% of total training steps.
The remaining 20\% of the training steps will keep the same mask and resume accuracy.

The number of weights to be pruned is $D=\alpha \sum(m^r\odot m^c)$.
The mask update procedure contains the following stages:
\ding{202} \textbf{Determine number of columns to prune}.~
Since the row sparsity pattern is fixed and identical for all blocks, we can calculate the number of column vectors to prune as $n_c=D/(\sum m^r/(pq))$ to match the scheduled death rate.
\ding{203} \textbf{Select small-magnitude column vector candidates}.~
Unpruned column vectors are sorted based on their $\ell$2-norm, $\|w\|_2$.
The smallest $n_c+\Delta m$ vectors form a pruning candidate pool.
$\Delta m$ is the selection margin (e.g., set to 2) to leave space for power optimization.
\ding{204} \textbf{Select low-power column vector candidates}.~
Among $n_c+\Delta m$ candidate vectors, we enumerate all $\binom{n_c+\Delta m}{n_c}$ combinations (up to a maximum combination in case there are too many candidates) and select the combination that minimizes the overall power consumption $P$.
We find the death mask $m_{\text{death}}^c$ where 1 represents newly pruned columns and update the column mask $m^c\gets m^c ~\&~ m^c_{\text{death}}$.

\begin{algorithm}[tb]
\caption{Power/Crosstalk-Aware Dynamic Sparse Training}
\label{alg:DSTAlgo}
\begin{algorithmic}[1]
\small 
\Require Loss function $\calL$, neural network $f_{b_{in}}(\cdot)$ with $b_{in}$-bit activation quantization, weight bitwidth $b_{w}$, input data $X$, target $y$, learning rate $\eta$, total steps $T$, steps per epoch $\Delta T$, epoch to stop prune\&grow $T_{end}$, pruning margin $\Delta m$, initial death rate $\alpha^0$.
\Ensure Converged parameter $W$ and sparsity mask $m$;

\State \textcolor{gray}{-------Crosstalk/Power-Minimized Initialization-------}
\State $s^r=\max(s,0.5)$, $m^{(l,r)}=\texttt{InterleavedOnes}(s^r)$ \Comment{Min-crosstalk $m^r$}
\State $s^c=s/s^r$, $m^{(l,c)}=\argmin_{m^{(l,c)}}P(m^{(l,c)})$ \Comment{Min-power $m^c$}
\For{$t \gets 1 \cdots T$}
\State $W^l\gets W^l \odot m^{(l,r)}\odot m^{(l,c)}$ \Comment{Inplace apply sparsity mask}
\State $W^l\gets W^l - \eta \nabla_{W^l}\calL\big(f_{b_{in}}(\calQ_{b_{w}}(W),X),y\big)$

\If{$t$ mod $\Delta T ==0$ and $t<T_{end}$}
\State $\alpha = \frac{\alpha^0}{2}(1+\cos(\frac{t\pi}{T_{end}}))$ \Comment{Schedule death rate}
\State \textcolor{gray}{-------Stage 1: Update Sparsity Mask with Pruning-------}
\State $D^{l}=[\alpha \sum(m^{(l,r)}\odot m^{(l,c)})]$ 
\State $n_c^{l}=D^{(l)}/(\sum m^{(l,r)}/(p^lq^l))$
\State Select ($n^l_c+\Delta m$) column vectors with smallest $\ell_2$-norm.
\State Further select the lowest power and least crosstalk column vectors.
\State $m^{(l,c)} \gets m^{(l,c)} ~\&~ m^{(l,c)}_{\text{death}}$
\State \textcolor{gray}{-------Stage 2: Update Sparsity Mask with Growth-------}
\State $n_c^l=(s^lp^lq^lrck_1k_2-\sum(m^{(l,r)}\odot m^{(l,c)}))/(\sum(m^{(l,r)})/(p^lq^l))$ \Comment{Number of columns to grow}
\State $m^{(l,c)} \gets m^{(l,c)} ~|~ m^{(l,c)}_{\text{grow}}$ \Comment{Similar procedure to select column vectors with large gradient norm and lowest power to grow}  
\EndIf
\EndFor
\end{algorithmic}
\end{algorithm}

\noindent\textbf{\underline{Power-Aware Growth Procedure}}.~
To maintain sparsity while exploring patterns, we grow (resume) roughly the same number of weights that were pruned.
The number of column vectors to be resumed $n_c$ is calculated based on the target sparsity $s$ and the number of nonzero elements per column, $n_c=(spqrck_1k_2-\sum(m^r\odot m^c))/(\sum(m^r)/(pq))$.
The column vector selection procedure is based on gradient magnitude for accuracy, i.e., $\|\frac{\partial \calL}{\partial w}\|_2$.
The same power minimization procedure applies to resume low-power column vectors.
At the end of the growth stage, we obtain a growth column mask $m^c_{\text{grow}}$, where 1 represents resumed columns.
We then update the sparsity mask, i.e., $m^c\gets m^c | m^c_{\text{grow}}$.

\section{Experimental Results}
\label{sec:ExperimentalResults}
\subsection{Experiment Setup}
\label{sec:ExpSetup}
\noindent\textbf{Dataset and Models.}~ 
We evaluate our method on a three-layer CNN (C64K3-C64K3-C64K3-Pool5-FC10) on Fashion-MNIST, VGG-8 on CIFAR-10, and ResNet-18 CIFAR-100 for image classification.

\noindent\textbf{Training Settings.}~
We pre-train CNN for 50 epochs with an Adam optimizer with a 2E-3 learning rate (lr), a cosine decay scheduler, 1E-4 weight decay, and data augmentation (random crop and flip) on Fashion-MNIST.
Other models are trained for 200 epochs with an SGD-momentum optimizer (lr of 0.02 for ResNet, 0.002 for VGG8).
We use learned stepsize quantization-aware training~\cite{NN_ICLR2020_Esser}.
we employ $b_w$=8-bit symmetric signed per-tensor quantization for weights and $b_{in}$=6-bit for activations.
For DST, we adopt an initial death rate of $\alpha^0$=0.5, $T_{end}$=80\% total training steps.
We update masks per epoch.
No noise-aware training is applied, which is orthogonal to our method.

\noindent\textbf{Architecture Settings}.~
We configure our architecture to have $R=4$ tiles with $C=4$ cores per tile.
Each PTC is of size $k_1=k_2=16$ working at clock frequency $f$=5 GHz.
We assume the same device cost as prior work~\cite{NP_HPCA2024_Zhu}.
For the MZI power splitter, we have two options: the one from foundry has 30 mW $P_{\pi}$ with 156.25 $\mu m$ in width and 550 $\mu m$ in length (Foundry-MZI); our optimized low-power MZI (LP-MZI) has a length of 115 $\mu m$ and width of $l_s+w_{PS}=9+6=15\; \mu m$ and a power profile shown in Fig.~\ref{fig:CrosstalkMZIPower} ($P_{\pi}= 15.02\;mW$).

\noindent\textbf{Evaluation Metrics}.~
We evaluate the total accelerator area ($A$), total energy $E_{\text{tot}}=\sum_l^L\sum_{i}^{p}\sum_{j}^{q} (P^l_{i,j}\cdot\texttt{Cyc}^l_{i,j}/f)$, calculated by accumulating each PTC's power over its execution runtime across all layers and all weight chunk, and
average power $P_{\text{avg}}=\frac{E_{\text{tot}}}{\texttt{Cyc}_{\text{tot}}/f}$.
We compare the area-energy efficiency (TOPS/W/mm$^2$) for efficiency evaluation.
To clarify, since a fine-grained row-column sparse model consumes the same cycle as a dense model, i.e., it still takes 1 cycle to map a $rk_1\times ck_2$ weight block onto our accelerator regardless of row/column sparsity, allowing us to use power-area product (PAP) to guide the optimization, which is equivalent to TOPS/W/mm$^2$ given the same speed (lower PAP means higher TOPS/W/mm$^2$).
Note that the memory latency of loading sparse/dense weights is often hidden by optimized SRAM design~\cite{NP_HPCA2024_Zhu}.
Hence, we do not show throughput/speed in our results.
We evaluate ideal accuracy and accuracy with crosstalk and random noises.
Since the last layer is sensitive to error and pruning, we protect the last linear layer by mapping the weights to non-adjacent columns of MZIs to eliminate crosstalk.
Note that we focus on the robustness and efficiency benefits from our circuit sparsity and light redistribution techniques on crossbar-style photonic tensor cores. 
Comparing the case-study architecture with other PTC designs or electronic digital accelerators is out of scope.

\subsection{Ablation Study}
We first explore different spaces to find optimal device/architecture settings and validate the effectiveness of our proposed techniques.
\subsubsection{Optimal Device Spacing}
\begin{table}[]
\caption{Optimal device spacing on a dense network ($s$=1) with high accuracy under crosstalk and noises ($\sim$1\% drop than ideal accuracy 91.4\%) and minimum power-area product (PAP). 
Average power $P_{\text{avg}}$ is evaluated on CNN-FashionMNIST
For a dense accelerator, the optimal settings are $l_s=9 \mu m$ and $l_g=5 \mu m$.}
\vspace{-3pt}
\label{tab:OptimalDeviceSpacing}
\resizebox{0.85\columnwidth}{!}{
\begin{tabular}{c|c|c|c|c|c}
\hline
$l_s$ ($\mu m$) & $l_g$ ($\mu m$) & Acc (\%)$\uparrow$ & $P_{\text{avg}}$(W)$\downarrow$ & $A$ (mm$^2$)$\downarrow$ & PAP $\downarrow$            \\ \hline
7  & 5  & 91.03     & 23.21    & 17.33   & 402.2          \\ \hline
8  & 5  & 91.11     & 22.06    & 17.81   & 393.0          \\ \hline
\rowcolor[HTML]{EFEFEF} 
9  & 5  & 91.10     & 20.58    & 18.30   & \textbf{376.6} \\ \hline
10 & 5  & 91.02     & 20.26    & 18.79   & 380.5          \\ \hline
11 & 5  & 91.00     & 19.70    & 19.27   & 379.8          \\ \hline
\end{tabular}
}
\vspace{-10pt}
\end{table}

Table~\ref{tab:OptimalDeviceSpacing} shows the trade-offs between spacing, area, power, and robustness across different MZI device spacing with a dense network ($s=1$).
Based on Fig.~\ref{fig:CrosstalkPowerArea}, we determine the most efficient arm spacing is $l_s$ is 9 $\mu m$, minimizing PAP.
To ensure <1\% accuracy drop, the minimum MZI horizontal gap $l_g$ is conservatively set to 5 $\mu m$.
In later experiments, we show sparsity and power gating enable further shrinking of the device spacing down to $l_g$=1$\mu m$ with superior crosstalk tolerance.

\subsubsection{Architecture Sharing Factor and Sparsity Granularity}
\begin{table}[]
\caption{Evaluate accuracy and inference average power on CNN-FashionMNIST with different sparsity, architecture sharing factor $r$ and $c$, and three sparsity.}
\label{tab:SparsityGranularity}
\resizebox{\columnwidth}{!}{
\begin{tabular}{cc|cc|cc|cc}
\hline
                       \multicolumn{2}{c|}{}  & \multicolumn{2}{c|}{Sparsity=0.8}            & \multicolumn{2}{c|}{Sparsity=0.6}            & \multicolumn{2}{c}{Sparsity=0.4}             \\ \hline
\multicolumn{1}{c|}{$r$} & $c$ & \multicolumn{1}{c|}{$P_{\text{avg}}$(W)$\downarrow$} & Acc (\%)$\uparrow$ & \multicolumn{1}{c|}{$P_{\text{avg}}$(W)$\downarrow$} & Acc (\%)$\uparrow$ & \multicolumn{1}{c|}{$P_{\text{avg}}$(W)$\downarrow$} & Acc (\%)$\uparrow$ \\ \hline
\multicolumn{1}{c|}{1} & 1 & \multicolumn{1}{c|}{17.94}        & 91.92    & \multicolumn{1}{c|}{17.22}        & 91.71    & \multicolumn{1}{c|}{17.99}        & 92.08    \\ \hline
\multicolumn{1}{c|}{2} & 2 & \multicolumn{1}{c|}{12.28}        & 91.86    & \multicolumn{1}{c|}{11.26}        & 91.73    & \multicolumn{1}{c|}{12.50}        & 91.69    \\ \hline
\rowcolor[HTML]{EFEFEF} 
\multicolumn{1}{c|}{4} & 4 & \multicolumn{1}{c|}{8.052}        & 91.78    & \multicolumn{1}{c|}{7.343}        & 91.76    & \multicolumn{1}{c|}{9.350}        & 91.85    \\ \hline
\end{tabular}
}
\end{table}

With a predefined PTC size $k_1 \times k_2=16\times 16$, we need to decide the optimal architecture sharing factor $r$ and $c$ to balance area, power, and accuracy.
Table~\ref{tab:SparsityGranularity} explores the impact of input/readout sharing factors ($r$, $c$) and sparsity.
A sharing factor $r=c=4$ offers the best power efficiency with minimum accuracy drop, corresponding to pruning/growth of length-64 weight row/column vectors.
With a larger accelerator scale, i.e., larger $R$ and $C$, we will keep the same sharing factor to maintain the pruning granularity.

\subsubsection{Light Redistribution and Power Gating}
\begin{figure}
    \centering
    \includegraphics[width=0.9\columnwidth]{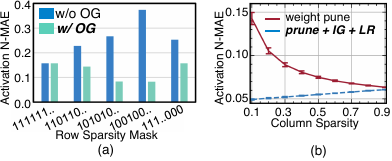}
    \vspace{-10pt}
    \caption{
    Thermal variation-induced activation error (N-MAE) on a 64-channel 3$\times$3 CONV layer.
    (a) Output TIA/ADC gating (w/ OG) with row sparsity masks with interleaved 1's can effectively reduce crosstalk-induced error.
    We use $l_s=9\mu m$ and $l_g=5\mu m$.
    (b) Input gating and light redistribution (IG+LR) can effectively suppress the error due to crosstalk and output noises.}
    \label{fig:RowSparsityOG}
\end{figure}

Figure~\ref{fig:RowSparsityOG}(a) shows different row sparsity patterns and their impacts on thermal variation-induced activation error.
Without TIA/ADC gating, sparse rows lead to even higher errors since zero elements still suffer from leakage and crosstalk errors.
With the proposed gating, the crosstalk-induced error will be eliminated, and noises will also be reduced.
Figure~\ref{fig:RowSparsityOG}(b) investigates the effectiveness of light gating and redistribution.
With lower sparsity (more zeros), the SNR will be largely increased with a significant fidelity boost.
In our \scatter system, we will enable OG+IG+LR together for the best thermal variation tolerance.

\begin{table*}[]
\caption{Evaluation of ideal accuracy, accuracy with thermal variation (w/ TV), resumed accuracy with input light gating (IG) + output TIA/ADC gating (OG) + light redistribution (LR), and single-image inference energy consumption.
CNN uses $s=0.3$, and VGG8/ResNet18 use $s=0.4$.
Device spacing settings and accelerator area are shown in the upper left corner of the table.
To clarify, the area adopts eoDAC, which is 0.704 mm$^2$ larger than the numbers shown in Fig.~\ref{fig:ProgressPowerAreaOpt}(\ding{204}\ding{205}).
}
\vspace{-5pt}
\label{tab:MainResults}
\resizebox{0.85\textwidth}{!}{
\begin{tabular}{|cc|cc|cc|cc|c|}
\hline
\multicolumn{2}{|c|}{}                                                                        & \multicolumn{2}{c|}{}                                                                                                              & \multicolumn{2}{c|}{}                                                                                                  & \multicolumn{2}{c|}{}                                                                                                  &                                                                         \\
\multicolumn{2}{|c|}{\multirow{-2}{*}{}}                                                      & \multicolumn{2}{c|}{\multirow{-2}{*}{\begin{tabular}[c]{@{}c@{}}$l_g$=1$\mu m$\\ Area=12.37 mm$^2$\end{tabular}}} & \multicolumn{2}{c|}{\multirow{-2}{*}{\begin{tabular}[c]{@{}c@{}}$l_g$=3$\mu m$\\ Area=13.44 mm$^2$\end{tabular}}}      & \multicolumn{2}{c|}{\multirow{-2}{*}{\begin{tabular}[c]{@{}c@{}}$l_g$=5$\mu m$\\ Area=14.20 mm$^2$\end{tabular}}}      &                                                                         \\ \cline{1-8}
\multicolumn{1}{|c|}{DensePTC}          & \begin{tabular}[c]{@{}c@{}}Ideal Acc\end{tabular} & \multicolumn{2}{c|}{Acc w/ TV}                                                                                                     & \multicolumn{2}{c|}{Acc w/ TV}                                                                                         & \multicolumn{2}{c|}{Acc w/ TV}                                                                                         & \multirow{-3}{*}{\begin{tabular}[c]{@{}c@{}}Energy\\ (mJ)\end{tabular}} \\ \hline
\multicolumn{1}{|c|}{CNN-FMNIST}        & 91.40                                               & \multicolumn{2}{c|}{84.00}                                                                                                         & \multicolumn{2}{c|}{89.10}                                                                                             & \multicolumn{2}{c|}{90.70}                                                                                             & 0.59                                                                    \\ \hline
\multicolumn{1}{|c|}{VGG8-CIFAR10}      & 88.02                                               & \multicolumn{2}{c|}{59.23}                                                                                                         & \multicolumn{2}{c|}{76.05}                                                                                             & \multicolumn{2}{c|}{81.54}                                                                                             & 3.17                                                                    \\ \hline
\multicolumn{1}{|c|}{ResNet18-CIFAR100} & 66.46                                               & \multicolumn{2}{c|}{44.12}                                                                                                         & \multicolumn{2}{c|}{57.84}                                                                                             & \multicolumn{2}{c|}{60.94}                                                                                             & 24.06                                                                   \\ \hline\hline
\multicolumn{1}{|c|}{\scatter}           & \begin{tabular}[c]{@{}c@{}}Ideal Acc\end{tabular} & \multicolumn{1}{c|}{Acc w/ TV}       & \cellcolor[HTML]{EFEFEF}\begin{tabular}[c]{@{}c@{}}Acc w/ TV\\ +IG+OG+LR\end{tabular}       & \multicolumn{1}{c|}{Acc w/ TV} & \cellcolor[HTML]{EFEFEF}\begin{tabular}[c]{@{}c@{}}Acc w/ TV\\ +IG+OG+LR\end{tabular} & \multicolumn{1}{c|}{Acc w/ TV} & \cellcolor[HTML]{EFEFEF}\begin{tabular}[c]{@{}c@{}}Acc w/ TV\\ +IG+OG+LR\end{tabular} & \begin{tabular}[c]{@{}c@{}}Energy\\ (mJ)\end{tabular}                   \\ \hline
\multicolumn{1}{|c|}{CNN-FMNIST}        & 91.56                                               & \multicolumn{1}{c|}{91.23}           & \cellcolor[HTML]{EFEFEF}91.26                                                               & \multicolumn{1}{c|}{91.24}     & \cellcolor[HTML]{EFEFEF}91.21                                                         & \multicolumn{1}{c|}{91.31}     & \cellcolor[HTML]{EFEFEF}91.30                                                         & 0.14                                                                    \\ \hline
\multicolumn{1}{|c|}{VGG8-CIFAR10}      & 85.64                                               & \multicolumn{1}{c|}{63.49}           & \cellcolor[HTML]{EFEFEF}82.04                                                               & \multicolumn{1}{c|}{72.78}     & \cellcolor[HTML]{EFEFEF}82.04                                                         & \multicolumn{1}{c|}{77.23}     & \cellcolor[HTML]{EFEFEF}82.24                                                         & 1.78                                                                    \\ \hline
\multicolumn{1}{|c|}{ResNet18-CIFAR100} & 59.18                                               & \multicolumn{1}{c|}{0.51}            & \cellcolor[HTML]{EFEFEF}57.40                                                               & \multicolumn{1}{c|}{0.86}      & \cellcolor[HTML]{EFEFEF}57.40                                                         & \multicolumn{1}{c|}{0.51}      & \cellcolor[HTML]{EFEFEF}57.46                                                         & 11.18                                                                   \\ \hline
\end{tabular}
}
\vspace{-5pt}
\end{table*}

\subsubsection{Progressive Power-Area Optimization}
\begin{figure}
    \centering
    \includegraphics[width=\columnwidth]{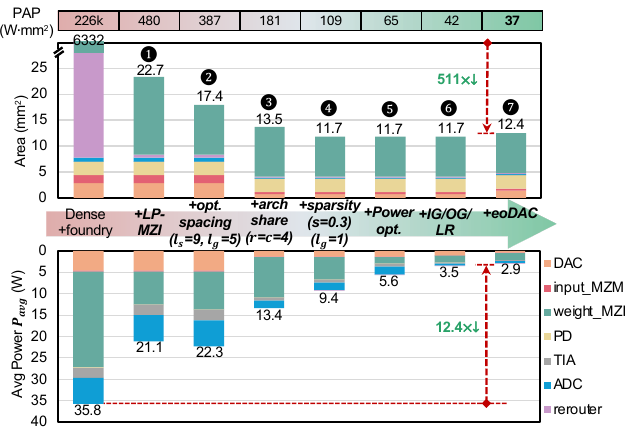}
    \vspace{-10pt}
    \caption{Significant power-area-product reduction can be achieved by progressively adding our proposed cross-layer optimization and algorithmic-circuit co-sparsity techniques.
    Detailed power/area breakdown has been presented.
    }
    \label{fig:ProgressPowerAreaOpt}
    \vspace{-10pt}
\end{figure}

Figure ~\ref{fig:ProgressPowerAreaOpt} illustrates the step-by-step impact of our optimizations towards orders-of-magnitude power/area reduction. 
The baseline is chosen to be a dense network with foundry MZI switches without architectural hardware sharing ($r$=$c$=1).
A conservative device spacing $l_g$=20 $\mu m$ is adopted to avoid thermal crosstalk issues.
\ding{202}~As we replace the foundry MZI with our \underline{compact low-power LP-MZI} device design, the chip area can be reduced by 279$\times$ with 41.1\% average power saving.
\ding{203}~We further squeeze the tensor core layout with our \underline{optimal device spacing}, i.e., $l_s$=9\;$\mu m$ and $l_g$=5\;$\mu m$, which gives a merely 5.7\% power penalty due to intra-MZI crosstalk but leads to 23.3\% area saving.
Note that this aggressive shrinking of gap $l_g$ causes severe inter-MZI crosstalk, leading to large accuracy degradation for a dense network.
\ding{204}~The \underline{architectural sharing} of input modulation and readout circuitry largely amortizes the DAC/ ADC cost, which further reduces the chip area by 39.9\% and saves 22\% power compared to dedicated DACs/ADCs for each PTC.
\ding{205}~In order to handle the thermal effect from shrinking the $l_g$, we add \underline{$s$=0.3 algorithm-circuit co-sparsity} to the accelerator. 
It allows us to turn off 70\% of the weight MZIs with 30\% power saving. 
With such interleaved row sparsity patterns, thermal crosstalk is mostly eliminated if output TIA/ADC gating (OG) is applied, which enables extremely narrow MZI spacing down to $l_g$=1 $\mu m$ with 13.3\% smaller chip area.
\ding{206}~During dynamic sparse training, we enable \underline{power-aware pruning/growth} to select low-power column masks.
\ding{207}~With \underline{input/output gating and light redistribution}, we not only suppress most of the thermal variations but also actively turn off the unused DACs/MZMs and TIAs/ADCs.
Also, power optimization helps to locate the least-power rerouter configurations.
\ding{208}~At the final step, we upgrade the traditional 6-bit eDAC with our optimized \underline{hybrid eoDAC} comprised of two 3-bit eDACs and a two-segment MZM.
We trade 2$\times$ the DAC area for 2.28$\times$ power reduction, which overall boosts the system power-area product by another 12\%.

\subsection{Main Results}
We compare the accuracy, power, and area on 2 settings and 3 benchmarks: (1) dense model and (2) \scatter with power-optimized sparsity in Table~\ref{tab:MainResults}.
Both settings adopt the best configurations from Fig.~\ref{fig:ProgressPowerAreaOpt}.

Models are evaluated under different $l_g$ with and without thermal variations. 
As the $l_g$ decreases from $5\; \mu m$ to $1\; \mu m$,  we can see a clear accuracy drop due to thermal variations for dense models.
With a row-column sparsity $s$=0.3$\sim$0.4, we observe some improvement on small benchmarks but much worse results on VGG8 and ResNet18.
\textbf{Key Insights:}
\ding{202}~Sparsity itself does not naturally boost the thermal robustness.
A sparse ResNet18 suffers from complete malfunction under crosstalk.
This is expected based on our previous analysis in Fig.~\ref{fig:RowSparsityOG}.
\ding{203}~Dense models degrade with smaller $l_g$ due to crosstalk, while \scatter resumes accuracy, when sparsity meets \emph{in-situ} power gating and light redistribution (IG+OG+LR).
\ding{204} Sparsity can enable extremely narrow MZI spacing $l_g$=1 $\mu m$ to save chip real estate by another 12.9\%. 
With input/output power gating, the single-image inference energy on three benchmarks can be reduced by an average of 52.9\%.
Our experiments demonstrate that \scatter's hardware/algorithm co-design significantly improves power efficiency and enables more compact photonic accelerators while maintaining thermal crosstalk robustness.
\vspace{-15pt}

\vspace{-5pt}
\section{Conclusion and Discussion}
\label{sec:Conclusion}
In this work, we introduce \scatter, \emph{the first} dynamically reconfigurable photonic tensor core architecture featuring cross-layer optimization for power, area, and thermal robustness.
Our \emph{in-situ} light redistribution and power gating enable fine-grained signal path control, facilitating algorithm-circuit sparsity co-exploration for significant power reduction and thermal variation suppression. 
Our power/crosstalk-aware dynamic sparse training framework automatically explores thermally robust, low-power sparsity masks tailored to \scatter hardware. 
We integrate synergistic optimization with customized compact low-power photonic devices, hybrid electrical-optical DACs, and optimal circuit/architecture design space exploration to maximize efficiency. 
Compared to dense photonic accelerators based on standard foundry devices, \scatter can save chip area by 511$\times$ and on-chip power consumption by 12.4$\times$, maintaining deployment accuracy even with significant thermal crosstalk and chip noise. 
This framework's dynamic reconfiguration and flexible signal path control establish a crucial design principle for next-generation reconfigurable photonic AI systems, pushing the boundaries of compute density and energy efficiency. 
Thermal crosstalk suppression via hardware/algorithm co-sparsity can be applied to other crossbar-type photonic tensor core designs, offering a generalizable and versatile co-design solution for reliable and efficient photonic AI computing systems.

{\small

\balance
\bibliographystyle{ACM-Reference-Format}
\bibliography{./ref/Top_sim,./ref/NN,./ref/NP,./ref/ALG, ./ref/addition, ./ref/IEEESettings, 
./ref/Defen}
}
\end{document}